\documentclass{ws-mpla}
\usepackage[super]{cite}
\usepackage{graphicx}
\begin{document}

\markboth{K.~Osetrin, A.~Filippov, E.~Osetrin}
{The space-time models with dust matter that admit separation of variables in Hamilton-Jacobi equations}

\catchline{}{}{}{}{}

\title{THE SPACE-TIME MODELS WITH DUST MATTER THAT ADMIT SEPARATION OF VARIABLES IN HAMILTON-JACOBI EQUATIONS OF A TEST PARTICLE}

\author{\footnotesize KONSTANTIN OSETRIN
}

\address{Theoretical Physics Department, Tomsk State Pedagogical University,\\
Tomsk, 634061, Russia\\
osetrin@tspu.edu.ru\\
and\\
National Research Tomsk State University,\\ 
Tomsk, 634050, Russia}

\author{ALTAIR FILIPPOV}

\address{Department of Informatics, Tomsk State Pedagogical University,\\
Tomsk, 634061, Russia\\
altair@tspu.edu.ru}

\author{\footnotesize EVGENY OSETRIN\footnote{zyxel@tspu.edu.ru
}}

\address{Theoretical Physics Department, Tomsk State Pedagogical University,\\
Tomsk, 634061, Russia\\
zyxel@tspu.edu.ru}

\maketitle

\pub{Received (Day Month Year)}{Revised (Day Month Year)}

\begin{abstract}
The characteristics of dust matter in space-time models, admitting the existence of privilege coordinate systems are given, where the single-particle 
Hamilton-Jacobi equation can be integrated by the method of complete separation of variables. The resulting functional form of the 4-velocity field 
and energy density of matter for all types of spaces under consideration is presented.

\end{abstract}

\ccode{04.20.Jb: include 11.30.Ly}


\section{Introduction}

At present, the basic constructive method for integrating geodesic equations in metric gravity theories is the method of complete separation of 
variables in the Hamilton-Jacobi equation for test particles. On the other hand, dust matter moving along geodesic lines of space-time is the standard 
model in the study of cosmological and astrophysical problems in the framework of metric theories of gravitation, including modified theories of 
gravity.

The Hamilton-Jacobi equation for test particles is as follows:
\begin{equation}
g^{ij}S_{,i}S_{,j}=m^2,
\qquad
i,j,k=0...3,
\label{eq0}
\end{equation} 
here $S$ -- the action of a test particle, $m$ -- mass of a 
particle. The spaces that admit of a "privileged" coordinate systems, where (\ref{eq0}) admit a complete separation of variables are called Stackel spaces (SS), see  [\refcite{Stackel}], [\refcite{1}]. The main results of Stackel spaces theory can be found in  [\refcite{2}],  [\refcite{3}], [\refcite{4}].

SS covariant condition is the presence of the so-called “complete set” of a commuting Killing vector and tensor fields, which satisfy some additional 
algebraic relations. Type of the SS metric tensor in privileged coordinate systems (where separation of variables is accepted) is determined up to a 
set of arbitrary functions where each function depends only on one variable. Types of SS differ on the number accepted in a complete set of commuting 
Killing vectors $Y^i_{(p)}$ ($p=1,N$)
 and the presence (absence) among the separated variables of the “wave” (null) coordinates. In total, there are seven types of 4- dimensional SS with Lorentz signature. The SS type is defined by a set of two numbers  $(N.N_0)$, where $N$ -- the number of commuting Killing vectors 
accepted by the space (the dimension of the Abelian group of space-time motions), and $N_0=N-rank|Y^i_{(p)}g_{ij}Y^j_{(q)}|$ -- the number of (null) variables in privileged coordinate 
systems (for 4-dimensional spaces of Lorentz signature $N=0...3$, $N_0=0,1$).

SS application in gravitation theories [\refcite{5}--\refcite{13}] is based on the fact that exact integrable models can be developed for these spaces. The majority of 
well-known exact solutions is classified as SS (Schwarzschild solutions, Kerr, Friedman, NUT, etc.). It is important to note that 
the other single-particle equations of motion - Klein-Gordon-Fock and Dirac, Weyl admit separation of variables only in SS. The same methods can be 
used to obtain solutions to the field equations in the theories of modified gravity [\refcite{14}], [\refcite{15}].

The energy-momentum tensor of dust matter is as follows:
\begin{equation} T_{ij}=\rho\, u_iu_j, 
\label{eq1}\end{equation} 
where $\rho$ -- energy density, $u_i$  -- field of matter velocity.

Implementation of the law of conservation is expected for the matter (the equations of the 
matter’s motion):
\begin{equation} \nabla^iT_{ij}=0\label{eq2}.\end{equation} 

The velocity vector of the matter corresponds to the norm condition (the space 
signature $(+,-,-,-)$):
\begin{equation}  u^iu_i=1.\label{eq3}\end{equation} 

The velocity vector of the matter is “separated” in a privileged coordinate system, i.e. 
corresponding covariant velocity components depend only on one variable:
\begin{equation} u_i=u_i(x^i). \end{equation} 
In the paper, the functional form of energy density and velocity components of dust matter in 
privileged coordinate systems is obtained for all types of SS. Privileged coordinate systems 
admit separation of variables in the equations  (\ref{eq0}), (\ref{eq3})
and the equations of the law of 
conservation of the energy-momentum  (\ref{eq2}) are carried out.

\section{Dust matter in Stackel spaces}

In privileged coordinate systems, variables (the metric is independent from them) are called 
“ignorable”. Thus, the geometric part of the gravitational equations, velocity components and 
energy density of the matter do not depend on the ignored variables. Nonignorable variables 
will be numbered by Greek indices  $\mu, \nu.$  The functions of a single variable will be supplied 
with the subscript which corresponds to the variable index, i.e.. $a_0=a_0(x^0)$, $b_1=b_1(x^1)$.

In the paper, the following notations will be used:
\begin{equation}
P=\ln\left |\frac{\rho^2}{\Delta^2 \det g^{ij}}\right |,
\end{equation} 
where $\Delta$ -- a conformal factor of the metrics (for some types of spaces $\Delta=1$).


\subsection{Stackel spaces of (3.0) type}

Stackel spaces of (3.0) type accept 3 commuting Killing vectors. In a privileged coordinate system, 
the metric depends only on one variable $x^0$:
\begin{equation} g^{ij}=\left(\begin{array}{cccc}1&0&0&0\\ 0&a_0&b_0&c_0\\ 0&b_0&d_0&e_0\\ 0&c_0&e_0&f_0 \end{array} \right)\end{equation} 
\[  \Delta=1, \qquad a_0, b_0, c_0, d_0, e_0, f_0 - \mbox{arbitrary functions of the variable } x^0. \]
The 4-velosity of matter in a privileged coordinate system has the following “separated” form:
\[ u_0=u_0(x^0),\qquad u_1=\alpha,\qquad u_2=\beta,\qquad u_3=\gamma, \qquad  \alpha, \beta, \gamma  - const.\]
The norm condition  (\ref{eq3})  provides the relation:
\begin{equation}
\alpha^2a_0+2\alpha\beta b_0+\beta^2d_0+2\alpha\gamma c_0+2\beta\gamma e_0+\gamma^2 f_0+u_0{}^2=1.
\end{equation}
The equations of motion  (\ref{eq2}) can be reduced to an equation for the $P$ function:
\[
u_0P_{,0}+(\alpha a_0+\beta b_0+\gamma c_0)P_{,1}+(\alpha b_0+\beta d_0+\gamma e_0)P_{,2}+\mbox{}
\]
\begin{equation}
\mbox{}+	(\alpha c_0+\beta e_0+\gamma f_0)P_{,3}+2u_0{}'=0.\label{eq303}
\end{equation}
Hence, we get two cases for velocity and energy density of the matter.

\subsubsection{Case (3.0 -- A). $ u_0\neq 0$.}
 \begin{equation} 
u_i=(u_0, \alpha,\beta,\gamma), 
\end{equation}
\begin{equation}
u_0=\sqrt{1-(\alpha^2a_0+2\alpha\beta b_0+\beta^2d_0+2\alpha\gamma c_0+2\beta\gamma e_0+\gamma^2 f_0)},
\end{equation}
\begin{equation}
 	\rho=\mbox{const}\frac{\sqrt{- \det g^{ij}}}{u_0}.
\end{equation} 

\subsubsection{Case (3.0 -- B). $ u_0= 0$.}
The function $\rho=\rho(x^0)$ is arbitrary, and we have the conditions:
\begin{equation}
u_i=(0, \alpha,\beta,\gamma), 
\qquad
\alpha^2a_0+2\alpha\beta b_0+\beta^2d_0+2\alpha\gamma c_0+2\beta\gamma e_0+\gamma^2 f_0=1.
\end{equation} 


\subsection{Stackel spaces of (3.1) type}

The space of (3.1) type admits 3 commuting Killing vectors. In a privileged coordinate system, 
the metric depends only on one variable $x^0$. The variable $x^0$ -- a null (wave) variable. The 
metric, which accepts a complete separation of variables of (3.1) type, can be written  in a privileged 
coordinate system as:
\begin{equation} g^{ij}=\left(\begin{array}{cccc} 0&1&a_0&b_0\\ 1&0&0&0\\ a_0&0&c_0&f_0\\ b_0&0&f_0&d_0 \end{array}\right)\end{equation} 

\[  \Delta=1,  \qquad a_0, b_0, c_0, d_0, f_0 - \mbox{arbitrary functions of } x^0.\]
The 4-velosity of matter is as follows:
\[ u_0=u_0(x^0),\qquad u_1=\alpha,\qquad u_2=\beta,\qquad u_3=\gamma, \qquad  \alpha, \beta, \gamma  - const.\]
The system of equations  (\ref{eq2})-(\ref{eq3})  can be reduced to two equations:
\begin{equation} \beta^2 c_0+2\beta\gamma f_0+\gamma^2 d_0+2(\alpha+\beta a_0+\gamma b_0)u_0=1,\quad \alpha^2+\beta^2+\gamma^2 \neq 0, \end{equation} 
\[ (\alpha+\beta a_0+\gamma b_0)P_{,0}+u_0P_{,1}+(a_0u_0+\beta c_0+\gamma f_0)P_{,2}+\]
	\begin{equation}  +(b_0u_0+\beta f_0+\gamma d_0)P_{,3}+2\beta a_0'+2\gamma b_0'=0,\label{eq313}\end{equation} 
Hence, for matter velocity and energy density, we obtain the following two cases.

\subsubsection{Case (3.1 -- A).  $\alpha+\beta a_0+\gamma b_0\neq 0$.}
  \begin{equation} u_0=\frac{1-(\beta^2 c_0+2\beta\gamma f_0+\gamma^2 d_0)}{2\,(\alpha+\beta a_0+\gamma b_0)},
\qquad
	\rho=\mbox{const}\frac{\sqrt{- \det g^{ij}}}{\alpha+\beta a_0+\gamma b_0}.\end{equation} 

\subsubsection{Case (3.1 -- B). $\alpha+\beta a_0+\gamma b_0 = 0$.}
Functions $u_0(x^0)$ and $\rho=\rho(x^0)$ remain arbitrary, with:
\begin{equation}
u_i=\left(u_0(x^0), \alpha,\beta,\gamma\right), 
\end{equation}
\begin{equation}
\alpha+\beta a_0+\gamma b_0 = 0
\qquad
 \beta^2 c_0+2\beta\gamma f_0+\gamma^2 d_0=1, \quad \beta^2+\gamma^2 \neq 0.
\end{equation}


\subsection{Stackel spaces of (2.0) type}

The space of this type admits two commuting Killing vectors. In a privileged coordinate 
system, the metric depends on two variables $x^0$ and $x^1$ :
\begin{equation} g^{ij}=\frac 1\Delta\left(\begin{array}{cccc}1&0&0&0\\ 0&{\epsilon}&0&0\\ 0&0&A&B\\ 0&0&B&C \end{array}\right)\label{metric20}\end{equation} 
\[ \Delta=t_0(x^0)+t_1(x^1),\quad A=a_0(x^0)+a_1(x^1),
\]
\[ B=b_0(x^0)+b_1(x^1),\quad C=c_0(x^0)+c_1(x^1), \quad \epsilon=\pm 1.\]
\[ u_0=u_0(x^0),\qquad u_1=u_1(x^1),\qquad u_2=\alpha,\qquad u_3=\beta,  \qquad  \alpha, \beta, \gamma  - const.\]
From the norm condition for velocity (\ref{eq3}) we have:
\begin{equation}
t_0={u_0}^2+\alpha^2 a_0+2\alpha\beta b_0+\beta^2 c_0+\gamma,\quad
 t_1=\epsilon{u_1}^2+\alpha^2 a_1+2\alpha\beta b_1+\beta^2 c_1-\gamma. \label{eq201}
\end{equation}
From the conservation law (\ref{eq2}) we obtain the equation for energy density:
\begin{equation}
u_0P_{,0}+{\epsilon}u_1P_{,1}+(\alpha A+\beta B)P_{,2}+(\alpha B+\beta C)P_{,3}+2u_0'+2{\epsilon}u_1'=0.
\end{equation}
For the matter velocity and energy density, we obtain the following cases.

\subsubsection{Case (2.0 -- A). $u_0u_1\neq 0$.}
 \begin{equation} 
u_i=(u_0, u_1,\alpha,\beta),
\end{equation}
\begin{equation}
u_0=\sqrt{t_0-\alpha^2 a_0-2\alpha\beta b_0-\beta^2 c_0-\gamma},
\end{equation}
\begin{equation}
u_1=\sqrt{\epsilon(t_1-\alpha^2 a_1-2\alpha\beta b_1-\beta^2 c_1+\gamma)},
\end{equation} 
\begin{equation}
\rho=F(X)\,\frac{\Delta\sqrt{-\det g^{ij}}}{u_0\,u_1},
\qquad
X=\int\frac{dx^0}{u_0}-\epsilon\int\frac{dx^1}{u_1},
\end{equation} 
where  $F(X)$ --- an arbitrary function of its argument.

\subsubsection{Case (2.0 -- B). $u_0=0$ or $u_1=0$.}
\[
u_0=0,\qquad u_1\neq 0, 
\qquad
t_0=\alpha^2 a_0+2\alpha\beta b_0+\beta^2 c_0+\gamma,
\]
\begin{equation}
\rho=F(x^0,X)\,\frac{\Delta\sqrt{-\det g^{ij}}}{u_1},
\qquad
X=\int\frac{dx^1}{u_1}.
\end{equation} 
\[
u_0\neq 0,\qquad u_1=0, \qquad
t_1=\alpha^2 a_1+2\alpha\beta b_1+\beta^2 c_1-\gamma,\]
\begin{equation}
\rho=F(x^1,X)\,\frac{\Delta\sqrt{-\det g^{ij}}}{u_0},
\qquad
X=\int\frac{dx^0}{u_0}.
\end{equation} 
Where  $F$ --- an arbitrary function of its arguments.

\subsubsection{Case (2.0 -- C). $u_0=u_1=0$.}
The function $\rho=\rho(x^0,x^1)$ remains arbitrary. The following conditions are satisfied:
\begin{equation}
u_i=(0,0,\alpha,\beta), \quad
t_0=\alpha^2 a_0+2\alpha\beta b_0+\beta^2 c_0+\gamma,\quad
t_1=\alpha^2 a_1+2\alpha\beta b_1+\beta^2 c_1-\gamma.
\end{equation}


\subsection{Stackel spaces of (2.1) type}
The space of this type admits two commuting Killing vectors. In a privileged coordinate 
system, the metric depends on two variables $x^0$ and $x^1$ . The variable $x^1$ -- null ("wave" type). The metric in a privileged coordinate system can be written as:

\begin{equation} g^{ij}=\frac 1\Delta\left(\begin{array}{cccc} 1&0&0&0\\ 0&0&f_1&1\\ 0&f_1&A&B\\ 0&1&B&C
\end{array}\right)\end{equation} 
\[ \Delta=t_0(x^0)+t_1(x^1),\ A=a_0(x^0)+a_1(x^1),\ B=b_0(x^0)+b_1(x^1),\ C=c_0(x^0)+c_1(x^1).\]
For 4-velosiyu we have:
\[ u_0=u_0(x^0),\qquad u_1=u_1(x^1),\qquad u_2=\alpha,\qquad u_3=\beta,  \qquad  \alpha, \beta  - const.\]
Separation of variables in the norm condition for velocity (\ref{eq3}) provides ($\gamma$ -- const):
\begin{equation} t_0={u_0}^2+\alpha^2 a_0+2\alpha\beta b_0+\beta^2 c_0+\gamma,
\end{equation}
\begin{equation}
 t_1=2(\alpha f_1+\beta)u_1+\alpha^2 a_1+2\alpha\beta b_1+\beta^2 c_1-\gamma.\label{eq211}
\end{equation} 
 From the conservation law (\ref{eq2}) we obtain the equation for energy density:
\begin{equation}
 u_0P_{,0}+(\alpha f_1+\beta ) P_{,1}+(\alpha A+\beta B+f_1 u_1)P_{,2}+(\alpha B+\beta C+u_1)P_{,3}+2\alpha f_1'+2{u_0}'=0.\label{eq212}
\end{equation} 
From the equations (\ref{eq211})--(\ref{eq212}) for velocity and energy density of dust matter, we obtain 
expressions through the functions of the metric of  four types.

\subsubsection{Case (2.1 -- A).  $u_0(\alpha f_1+\beta)\neq 0$.}

 \begin{equation} 	 u_0=\sqrt{t_0-\alpha^2 a_0-2\alpha\beta b_0-\beta^2 c_0-\gamma},
\end{equation}
\begin{equation}
u_1=	 \frac{t_1-\alpha^2 a_1-2\alpha\beta b_1-\beta^2 c_1+\gamma}{2\,(\alpha f_1+\beta)},
\end{equation} 
\begin{equation} \rho=F(X)\,\frac{\Delta\sqrt{-\det g^{ij}}}{u_0(\alpha f_1+\beta)},
\qquad 
X=\int\frac{dx^0}{u_0}- \int\frac{dx^1}{\alpha f_1+\beta},
\end{equation} 
where $F$ --- an arbitrary function of its argument.

\subsubsection{Case (2.1 -- B).    $u_0=0$, $\alpha f_1+\beta\neq 0$.}

\begin{equation} 
t_0=\alpha^2 a_0+2\alpha\beta b_0+\beta^2 c_0+\gamma,\qquad
u_1=	 \frac{t_1-\alpha^2 a_1-2\alpha\beta b_1-\beta^2 c_1+\gamma}{2\,(\alpha f_1+\beta)},
\end{equation} 
\begin{equation}
 \rho=F(x^0, X)\,\frac{\Delta\sqrt{-\det g^{ij}}}{(\alpha f_1+\beta)},
\qquad 
X= \int\frac{dx^1}{\alpha f_1+\beta}.
\end{equation} 

\subsubsection{Case (2.1 -- C).   $\alpha f_1+\beta=0$, $u_0\neq 0$.}
The function $u_1(x^1)$ remains arbitrary.
\begin{equation}
u_0=\sqrt{t_0-\alpha^2 a_0-2\alpha\beta b_0-\beta^2 c_0-\gamma},
\end{equation}
\begin{equation} 
 t_1=\alpha^2 a_1+2\alpha\beta b_1+\beta^2 c_1-\gamma,
\end{equation}
\begin{equation}
\rho=F(x^1,X)\,\frac{\Delta\sqrt{-\det g^{ij}}}{u_0},
\qquad 
X=\int\frac{dx^0}{u_0}.
\end{equation} 

\subsubsection{Case (2.1 -- D).   $u_0=0$, $\alpha f_1+\beta=0$.}
The functions $u_1(x^1)$ and $\rho=\rho(x^0,x^1)$ remain arbitrary. We have the conditions:
 \begin{equation}
t_0=\alpha^2 a_0+2\alpha\beta b_0+\beta^2 c_0+\gamma, \qquad 
 t_1=\alpha^2 a_1+2\alpha\beta b_1+\beta^2 c_1-\gamma.
\end{equation}


\subsection{Stackel spaces of (1.0) type}
The space of this type admits one Killing vector. In a privileged coordinate system, the 
metric depends on three variables $x^1$,  $x^2$  and  $x^3$:
\begin{equation} g^{ij}=\frac 1\Delta\left(\begin{array}{cccc}\Omega &0&0&0\\ 0&V^1&0&0\\ 0&0&V^2&0\\ 0&0&0&V^3\end{array}\right)\end{equation} 
\[
\Delta=\sigma_\mu(x^\mu) V^\mu,\qquad
 \Omega=\omega_\nu(x^\nu) V^\nu, \qquad 
\mu,\,\nu=1...3.
\]
\[ V^1=t_2(x^2)-t_3(x^3),\quad V^2=t_3(x^3)-t_1(x^1),\quad V^3=t_1(x^1)-t_2(x^2),\]
\[ u_i=(\alpha,u_1(x^1),u_2(x^2),u_3(x^3)),\qquad \alpha=const.\]
The system of equations  (\ref{eq2})--(\ref{eq3}) will be:
\begin{equation} \Omega \alpha^2 + V^\mu {u_\mu}^2=\Delta,\label{eq101}\end{equation} 
\begin{equation}  \alpha \Omega P_{,0}+ V^\mu(u_\mu P_{,\mu}+2u_\mu')=0. \label{eq101-1}\end{equation} 
From the relation  (\ref{eq101}) and  (\ref{eq101-1}) we obtain the following cases.

\subsubsection{Case (1.0 -- A).   $u_1\,u_2\,u_3\neq 0$.}
\begin{equation}
u_\mu=\sqrt{\sigma_\mu-\alpha^2\omega_\mu+\beta t_\mu+\gamma},  \label{eq101-2}
\qquad
\mbox{$\beta,\gamma$ -- const}.
\end{equation} 
For energy density, from the equation  (\ref{eq101-1}) , we obtain the expression through the metric 
functions:
\begin{equation} 
\rho=F(X,Y)\,\frac{\Delta\sqrt{-\det g^{ij}}}{u_1\,u_2\,u_3},
\qquad
X=\sum_\mu  \int\frac{t_\mu}{u_\mu}\, dx^\mu ,
\qquad
Y= \sum_\mu \int\frac{dx^\mu}{u_\mu} ,
\end{equation} 
where $F(X,Y)$ --- an arbitrary function of its arguments.

\subsubsection{Case (1.0 -- B).   $u_1\,u_2\,u_3=0$.}
In case when some of the velocity components become zero (for example with the index  $\nu$), we 
obtain:
\begin{equation} 
u_\nu=0,
\qquad
\sigma_\nu=\alpha^2\omega_\nu-\beta t_\nu-\gamma,
\end{equation}
\begin{equation}
\rho=F(x^\nu,X,Y)\,\Delta\sqrt{-\det g^{ij}}/\prod_{\mu\neq \nu} u_\mu,
\end{equation}
\begin{equation}
X=\sum_{\mu\neq \nu} \int\frac{t_\mu}{u_\mu}\, dx^\mu ,
\qquad
Y= \sum_{\mu\neq \nu} \int\frac{dx^\mu}{u_\mu} ,
\end{equation}
\subsubsection{Case (1.0 -- C).  $u_1=u_2=u_3=0$.}
The function $\rho=\rho(x^1,x^2,x^3)$ remains arbitrary, with:
\begin{equation} 
u_i=(\alpha,0,0,0),
\qquad
\sigma_\mu=\alpha^2\omega_\mu-\beta t_\mu-\gamma,
\qquad \mu=1,2,3.
\end{equation} 
 

\subsection{Stackel spaces of (1.1) type}
The spaces of this type admits one Killing vector. In a privileged coordinate system, the 
metric depends on three variables $x^1$,  $x^2$ and $x^3$ . The variable $x^1$ -- is null ( “wave” type). The metric 
in a privileged coordinate system can be written as:
\begin{equation} g^{ij}=\frac 1\Delta\left(\begin{array}{cccc}\Omega&V^1&0&0\\ V^1&0&0&0\\ 0&0&V^2&0\\ 0&0&0&V^3\end{array}\right)\end{equation} 
\[ V^1=t_2(x^2)-t_3(x^3),\quad V^2=t_3(x^3)-t_1(x^1),\quad V^3=t_1(x^1)-t_2(x^2).\]
\[ \Delta=\phi_\mu (x^\mu)V^\mu,
\qquad 
\Omega=\omega_\mu(x^\mu) V^\mu,
\qquad
\mu,\nu=1...3,
\]
\begin{equation} u=(\alpha,u_1(x^1),u_2(x^2),u_3(x^3)),\qquad \alpha,\beta, \gamma - const.\end{equation} 
The system of equations  (\ref{eq2})-(\ref{eq3})  can be reduced to two equations:
\begin{equation} V^1(2\alpha u_1+\alpha^2\omega_1-\phi_1)+V^2(u_2{}^2+\alpha^2\omega_2-\phi_2)+
	V^3(u_3{}^2+\alpha^2\omega_3-\phi_3)=0,\label{eq111}\end{equation} 
\begin{equation} (V^1u_1+\alpha\Omega)P_{,0}+\alpha V^1P_{,1}+u_2V^2P_{,2}+u_3V^3P_{,3}+2V^2u_2'+2V^3u_3'=0.\label{eq112}\end{equation} 
We get the following cases of the relations for energy density and velocity components of 
the matter.

\subsubsection{Case (1.1 -- A).  $\alpha\, u_2 u_3\neq 0$.}
\begin{equation}  u_1=\frac 1{2\alpha}(\beta t_1+\gamma -\alpha^2\omega_1+\phi_1),
\end{equation} 
\begin{equation} 
u_2=\sqrt{\beta t_2+\gamma-\alpha^2\omega_2+\phi_2},
\qquad
u_3=\sqrt{\beta t_3+\gamma-\alpha^2\omega_3+\phi_3}.
\label{eq113}
\end{equation} 
\begin{equation} 
\rho=F(X,Y)\,\frac{\Delta\sqrt{-\det g^{ij}}}{u_2\,u_3},
\end{equation} 
\begin{equation} 
X=-\frac{1}{\alpha}\int t_1\, dx^1  +\int\frac{t_2}{u_2}\,dx^2+\int\frac{t_3}{u_3}\,dx^3 ,
\quad
Y= \frac{x^1}{\alpha}+\int\frac{dx^2}{u_2}+\int\frac{dx^3}{u_3},
\end{equation} 
where  $F(X,Y)$ --- an arbitrary function of its arguments.

\subsubsection{Case (1.1 -- B).    $u_2u_3=0$.}
In the case when some of the components of velocity  $u_2$ or $u_3$ become zero, the rest of the 
components are determined by the relations  (\ref{eq113}). For energy density, we have:
\begin{equation}
u_2=0, \qquad u_3\neq 0,\qquad
\rho=F(X,Y)\,\frac{\Delta\sqrt{-\det g^{ij}}}{u_3},
\end{equation} 
\begin{equation} 
X=-\frac{1}{\alpha}\int t_1\, dx^1 +\int\frac{t_3}{u_3}\,dx^3 ,
\qquad
Y= \frac{x^1}{\alpha}+\int\frac{dx^3}{u_3}.
\end{equation} 
\begin{equation}
u_2\neq 0,\qquad u_3= 0,\qquad
\rho=F(X,Y)\,\frac{\Delta\sqrt{-\det g^{ij}}}{u_2},
\end{equation} 
\begin{equation} 
X=-\frac{1}{\alpha}\int t_1\, dx^1 +\int\frac{t_2}{u_2}\,dx^2 ,
\quad
Y= \frac{x^1}{\alpha}+\int\frac{dx^2}{u_2}.
\end{equation} 

\subsubsection{Case (1.1 -- C).   $u_0=\alpha=0$,\quad $u_2\,u_3\neq 0$. }
\begin{equation}
\phi_1=-\beta t_1-\gamma,
\qquad
u_1=pt_1+q,
\end{equation} 
\begin{equation} 
u_2=\sqrt{\beta t_2+\gamma+\phi_2},
\qquad
u_3=\sqrt{\beta t_3+\gamma+\phi_3},
\qquad
 \mbox{$p$, $q$ -- const,}
\end{equation} 
\begin{equation}
\rho=F(x^1,X,Y)\,\frac{\Delta\sqrt{-\det g^{ij}}}{u_2\,u_3},
\end{equation} 
\begin{equation} 
 X=\int\frac{t_2}{u_2}\,dx^2+\int\frac{t_3}{u_3}\,dx^3 ,
\qquad
 Y= \int\frac{dx^2}{u_2}+\int\frac{dx^3}{u_3},
\end{equation} 
 where $F$  -- an arbitrary function of its arguments.

\subsubsection{Case (1.1 -- D).    $\alpha=u_2=u_3=0$.}
The functions $\rho=\rho(x^1,x^2,x^3)$ and $u_1(x^1)$ remain arbitrary, and from   (\ref{eq111})  it follows that: 
\begin{equation}
 \phi_\mu=-\beta t_\mu-\gamma,
\qquad 
\mu=1,2,3.
\end{equation} 


\subsection{Stackel spaces of (0.0) type}
In a privileged coordinate system, the metric (0.0) type depends on all variables:
\begin{equation} g^{ij}=\frac 1\Delta\left(\begin{array}{cccc}V^0 &0&0&0\\ 0&V^1&0&0\\ 0&0&V^2&0\\ 0&0&0&V^3 \end{array}\right)\end{equation} 
\[ \Delta=\phi_i (x^i)V^i, \qquad i=0...3,\]
\[ 
V^0=a_1(b_2-b_3)+a_2(-b_1+b_3)+a_3(b_1-b_2),
\quad 
V^1= a_0(-b_2+b_3)+a_2(b_0-b_3)+a_3(-b_0+b_2),
\]
\[
V^2=a_0(b_1-b_3)+a_1(-b_0+b_3)+a_3(b_0-b_1),
\quad
 V^3=a_0(-b_1+b_2)+a_1(b_0-b_2)+a_2(-b_0+b_1).
\]
Velocity of the matter has a “separated” form:
\[ u_i=(u_0(x^0),u_1(x^1),u_2(x^2),u_3(x^3)).\]
The norm condition and the equation of matter motion are as follows:
\begin{equation} V^iu_i{}^2=\Delta,\end{equation} 
\begin{equation} V^i(u_iP_{,i}+2u_i')=0,\end{equation} 
Hence, we obtain the following cases for the components of velocity and energy density of 
the matter.

\subsubsection{Case (0.0 -- A).    $u_0\,u_1\,u_2\,u_3\neq 0$.}
\begin{equation}
u_i=\sqrt{\phi_i +\alpha a_i+\beta b_i+\gamma},
\qquad
\mbox{$\alpha$, $\beta$, $\gamma$ -- const,}
\end{equation} 
\begin{equation} 
\rho=F(X,Y,Z)\,\frac{\Delta\sqrt{-\det g^{ij}}}{u_0\,u_1\,u_2\,u_3},
\end{equation} 
\begin{equation} 
X=\sum_i \int\frac{dx^i}{u_i},
\qquad
Y=\sum_i \int\frac{a_i}{u_i}\,dx^i,
\qquad
Z=\sum_i \int\frac{b_i}{u_i}\,dx^i,
\end{equation} 
where $F$ --- an arbitrary function of its arguments.

\subsubsection{Case (0.0 -- B).   $u_0\,u_1\,u_2\,u_3= 0$.}
In case when some of the velocity components, for example with  $k$ index, become zero, we have:
\begin{equation} 
u_k=0,
\qquad 
\phi_k =-\alpha a_k-\beta b_k-\gamma,
\end{equation} 
\begin{equation}
\rho=F(X,Y,Z)\,\Delta\sqrt{-\det g^{ij}}/ \prod_{i\neq k}{u_i},
\end{equation} 
\begin{equation} 
X=\sum_{i\neq k} \int\frac{dx^i}{u_i},
\qquad
Y=\sum_{i\neq k} \int\frac{a_i}{u_i}\,dx^i,
\qquad
Z=\sum_{i\neq k} \int\frac{b_i}{u_i}\,dx^i.
\end{equation}

\section{Stackel spaces of (2.1) type with dust matter and the cosmological constant in Einstein's 
gravity theory}

As an example, we consider the specific metric theory of gravity -- the General Relativity. We will find out a solution to Einstein field equations for SS of (2.1) type with 
dust matter and the cosmological constant:
\begin{equation}
R_{ij}-\frac{1}{2}g_{ij}R=\Lambda\, g_{ij}+\rho\, u_i u_j ,
\end{equation} 
where $\Lambda$ -- the cosmological constant, $\rho$ -- energy density of dust matter, $u_i$ -- the 4-velocity 
matter.

Norm conditions for the 4-velocity are carried out:
\begin{equation}
g^{ij}u_iu_j=1.
\end{equation} 
For the 4-velocity matter component, a privileged coordinate system involves a “separated”
form, i.e.  $u_i=u_i(x^i)$. 

In this case, a solution to Einstein's equations has the following two cases.

\subsection{Case A. $f_1\neq const$.}

The metric has the following form:
\begin{equation} g^{ij}=\frac 1{{u_0(x^0)}^2}\left(\begin{array}{cccc}1&0&0&0\\ 0&0&f_1(x^1)&1\\ 0&f_1(x^1)&a_0(x^0)+a_1(x^1)&b_0(x^0)\\ 0&1&b_0(x^0)&c_0(x^0) \end{array}\right),\end{equation} 
The functions $a_0(x^0)$, $b_0(x^0)$, $c_0(x^0)$, $f_1(x^1)$, $a_1(x^1)$ are given by:
\begin{equation} a_0=\lambda q_0,\ \ b_0=\mu,\ \ c_0=\sigma q_0+\nu,\ \ \lambda,\mu,\nu,\sigma - const,\end{equation} 
\begin{equation}{f_1'}^2=\lambda+\sigma {f_1}^2,\ \ 
\lambda^2+\sigma^2\neq 0, \quad \lambda\sigma=0,\end{equation} 
\begin{equation} a_1=2\mu f_1-\nu {f_1}^2,\end{equation} 
The functions $u_0(x^0)$  and $q_0(x^0)$ are determined by the following  equations:
\begin{equation} u_0''=\frac\Lambda{2}{u_0}^3+\frac{3u_0}{8q_0}+\frac{{u_0'}^2}{2u_0},\end{equation} 
\begin{equation} q_0''=2+\frac{3{q_0'}^2}{2q_0}-\frac{2q_0'u_0'}{u_0}.
\end{equation} 
For energy density and velocity of the matter, we have:
\begin{equation} u_i=\left(u_0(x^0),0,0,0\right),
\qquad
\rho=-\Lambda+\frac{3{u_0'}^2}{{u_0}^4}-\frac{u_0+4q_0'u_0'}{4q_0{u_0}^3}. \end{equation} 
Weyl tensor can not become zero, i.e., this space cannot be conformally flat.

\subsection{Case B. $f_1=0$.}
The metric takes the following form:
\begin{equation} g^{ij}=\frac 1{{u_0}^2}\left(\begin{array}{cccc} 1&0&0&0\\ 0&0&0&1\\ 0&0&a_1&b_0\\ 0&1&b_0&c_0\end{array}\right),\end{equation} 
The functions $a_1(x^1)$, $c_0(x^0)$, $u_0(x^0)$ are defined by the following equations:
\begin{equation}	{u_0'}^2=\frac\Lambda{3}{u_0}^4+\frac\lambda{3}{u_0},\quad \lambda - const,\end{equation} 
\begin{equation} {a_1'}^2=\kappa{a_1}^3-2\mu{a_1}^2,\qquad c_0''+2c_0'\frac{u_0'}{u_0}=-\mu, \qquad b_0=\nu,\qquad \kappa,\mu,\nu - const.\end{equation} 
For energy density and the matter velocity, we have:
\begin{equation} u_i=(u_0,0,0,0),
\qquad
\rho=\frac\lambda{{u_0}^3},
\end{equation} 
Weyl tensor can not become zero, i.e., this space cannot be conformally flat.

\section*{Acknowledgments}

In the paper we have considered the functional form of the energy-momentum tensor of dust 
matter (energy density and velocity) for all types of space-time, which admit integration of 
geodesic equations in the form of the Hamilton-Jacobi by method of a complete separation of 
variables. In privileged coordinate systems (separation of variables is admitted), energy 
density of the matter and velocity are determined via the functions in the space-time metric.

These results can be used to construct exact models for different metric theories of gravity, 
as well as for comparison with similar models in Einstein's theory and theories of modified 
gravity  [\refcite{14}], [\refcite{15}]. 

Solutions to Einstein's equations for the models with dust matter and the cosmological 
constant in a privileged coordinate system have been obtained for SS of (2.1) type.

Research was supported by Russian Ministry of Education and Science under contract 3.867.2014/K. 

Osetrin~K.E.  is grateful to the grant for LRSS, project 88.2014.2

%
%
%
%
\end{document}